\documentclass[twocolumn]{aastex61}
\usepackage{epsfig}

\newcommand{\lsim}{\hbox{\rlap{$^<$}$_\sim$}}

\begin{document}
\title{Universal Peaks Ratio In The Spectral Energy Density \\ 
     Of Double Hump Blazars, Gamma Ray Bursts, And Microquasars ?} 

 \author{Shlomo Dado}
\affiliation{Physics Department, Technion, Haifa 32000, Israel}
\author{Arnon Dar}
\affiliation{Physics Department, Technion, Haifa 32000, Israel}

\begin{abstract}

The peak frequencies of the two broad humps evident in the spectral 
energy density of blazars (SED) are time dependent and vary a lot 
between different blazars. However, their ratio in most blazars, appears 
to be almost universal and equal to $\!m_e c^2/4(1\!+\!z) \epsilon_p$ to 
a good approximation, where $m_e$ is the electron mass, $\epsilon_p$ is 
the peak energy of the cosmic microwave background radiation, and $z$ is 
the redshift of the blazar. We discuss a possible origin of such a 
universal ratio in blazars, gamma ray bursts (GRBs).
We point out a possible connection between the knee 
in the energy spectrum of cosmic ray electrons and the maximal peak 
energies of the two broad humps in the SED of high-energy peaked blazars 
and GRBs. We also point out  that a universal peaks ratio in double hump blazars
which belong to different classes in the BL Lac sequence,
may simply reflect different viewing angles of otherwise similar blazars. 
\end{abstract}


\section{Introduction}

Blazars form a subclass of active galactic nuclei (AGN) where the mass 
accreting central massive black hole fires a highly relativistic jet of 
plasmoids in a direction very close to that of Earth (Blandford \& Rees 1978; 
Urry \& Padovani 1995; Ulrich et al. 1997). As such, blazars are very 
luminous and violently variable over a large range of frequencies $\nu$ from 
radio to TeV gamma rays. A two hump structure in the broad band spectral 
energy density (SED)  $\nu\,F_\nu $ of blazars is evident when it is plotted 
as a function of $log\,\nu$ (Fossati et al. 1998). The low energy hump has a 
peak value anywhere in the radio to X-ray band, while the high energy hump 
has a peak in the MeV - TeV gamma ray band. Although the origin of these two 
humps has not been established beyond doubt, it is widely believed that the 
first hump is synchrotron radiation emitted by energetic electrons within the 
jet (Konigel 1981; Urry \& Mushotzky 1982) while the second hump is produced 
by inverse Compton scattering (ICS) of this synchrotron radiation by the 
relativistic electrons in the jet -the so called Synchrotron Self Compton 
(SSC) mechanism (Jones et al. 1974a,b)- or external photons from the 
accretion disk, and/or a dusty torus, and/or a broad line region, and/or 
galactic and extragalactic background radiations, the so called External 
Compton (EC) mechanisms (e.g., Marscher \& Gear 1985; Dermer et al. 1992; 
Dermer and Schlickeiser 1993; Sikora et al. 1994; Ghisellini \& Madau 1996); 
Hartman et al. 2001);  Krawczynski et al. 2001; Sikora et al. 2001; Sokolov 
\& Marsher 2005; Albert et al. 2008; Abdo et al. 2014; Yang et al. 2017a,b). 
Both the SSC and EC models of high energy emission had considerable success 
in fitting the observed broad band SED of many blazars. They have been used 
also to explain the recent discoveries of TeV gamma ray emission from the 
gamma ray burst (GRB) 190114C (Acciari et al. 2019a, Ajello et al. 2020) and 
from the terminal lobes of the bipolar jet of the Galactic microquasar SS 433 
(Abeysekara et al. 2018; Xing et al. 2019). However, the claimed success of 
both the SSC and EC models was based on posteriori fits to observational 
data, which involved many adjustable parameters and free choices, rather than 
on falsifiable predictions.

In this letter we show that although the peak frequencies of the 
two broad humps evident in the broad band SED 
of blazars vary between and within the different blazar classes, 
and depend on viewing angle and epoch, 
their ratio appears to be almost universal. We demonstrate 
that for a selected sample of blazars with a simultaneous, well sampled
broad band SED between radio and TeV energies. These blazars 
include both flat spectrum radio quasars (FSRQs) with a low
$\nu_{p1}$ (below $10^{14}$ Hz), and  the so called 
BL Lac sequence: LBL, IBL, HBL, and EHBL, with $\nu_{p1}$ values,  
low ($<\!10^{14}$ Hz), intermediate 
(between $10^{14}$ and $10^{15}$ Hz), high (between $10^{15}$ and 
$10^{17}$ Hz), and extremely high (above $10^{17}$ Hz), respectively
(Acciari et al. 2019b, and references therein).
To a good approximation, this universal ratio satisfies
\begin{equation}
(1\!+\!z)\nu_{p2}/\nu_{p1}\! \approx\!m_e c^2/4\,\epsilon_p(CMB)) 
\approx 1.88\times 10^8, 
\end{equation}
where $m_e c^2/2$ is the photon energy in the electron rest frame 
around which inverse Compton scattering changes from the Thomson 
regime to the Klein Nishina regime, and 
$(1\,+\,z)\epsilon_p\,=\,(1\,+\,z)0.68\times 10^{-3}$ eV is the peak 
energy of the cosmic microwave background (CMB) radiation (Fixsen 
2009) at the observed redshift $z$ of the blazar. 

We also show that the above universal ratio is expected if both humps 
are produced by ICS of isotropic distribution of external photons.
The low energy hump 
is produced by ICS of CMB photons in the Thomson regime by the inert 
electrons in the jet of plasmoids ("cannonballs") which are ejected 
with a high relativistic bulk motion in mass accretion episodes onto 
the central compact object. The high energy hump by ICS in the 
Klein-Nishina  regime of harder radiation fields encountered along the 
jet trajectory, by the ambient electrons of the surrounding medium 
which  are scattered by the plasmoids to cosmic ray (CR) energies.
The seed photons can be blazar radiations such as UV light and  
x rays from the accretion disk, and gamma rays from 
hadronic interaction in the broad line region of flat spectrum radio 
quasars (Dar and Laor 1997), but more likely, photons of the
nearly isotropic galactic and extragalactic background  
radiations along the jet trajectory, or synchrotron radiation emitted 
by Fermi accelerated electrons within the plasmoids of  blazar jets 
(e.g., Longair 2011 and references therein).

Finally, for completeness, we examine whether the  
discoveries of TeV gamma rays from  GRB 190114C (Acciari et al.
2019a, Ajello et al. 2020) and the Galactic microquasar SS 433
(Abeysekara et al. 2018; Xing et al. 2019) suggest a double  
hump broad band SED of GRBs (Dado \& Dar 2005) and microquasaes
with the nearly universal peak ratio observed in blazars. 
 
\section{Origin Of Blazar's Double Humps SED} 
Mass accretion episodes onto the central massive black hole in 
blazars launch highly relativistic jets of plasmoids of ordinary 
matter with an initial bulk motion Lorentz factor $\gamma(0)\!\gg\! 1$. 
These highly relativistic plasmoids (cannonballs) slow down mainly by 
gathering and scattering of the ionized nuclei and free electrons in 
front of them (e.g., Dar \& De R\'ujula 2008).  Consequently, 
initially the highly relativistic plasmoids contain two populations 
of electrons: the inert electrons of the plasmoids, and the external 
electrons which were swept in with a Lorentz factor $\gamma$ in the 
plasmoids rest frame and scattered within them by internal magnetic 
fields.
 
When a jet of plasmoids with a large bulk motion Lorentz factor 
$\gamma\!\gg\!1$ propagates through a radiation field, they 
produce, through ICS of ambient photons, a narrow beam of photons 
along the jet direction of motion. As long as the initial energy 
of the incident  photons in the electron rest frame is much 
smaller than $m_e c^2$ (the Thomson regime), the maximal energy that 
ambient photons with initial energy $h\nu_0$ acquire through ICS by 
electrons at rest within the plasmoid, is in head-on collisions 
in which they are scattered backward, i.e., in the direction of motion 
of the plasmoid. This energy in the blazar rest frame is  
\begin{equation}
h\nu_{max}\!\approx\! 4\gamma^2 h\nu_0 \,.
\end{equation}
The peak  energy of up scattered CMB photons by a highly
relativistic plasmoid  through 
ICS in the Thomson regime, which are viewed from a small angle 
$\theta\!\ll\!1$ relative to its direction of motion, is given by
\begin{equation}
E_{p1}\!\approx 2\! h\nu_{p1}\!\approx\!\gamma\,\delta \epsilon_p(CMB)\\,.
\end{equation}  
where $\delta\!=\!1/\gamma(1\!-\beta\,cos\theta)$ is the Doppler 
factor, and $\epsilon_p(CMB)$ is the peak energy of the CMB photons. 
Note that $E_{p1}$ is independent of the blazar redshift. This is 
because the blue shift of the CMB temperature at the blazar 
emission time is equal to the redshift of the emitted radiation on 
its way to Earth.

The interstellar medium (ISM) electrons which are scattered forward 
by the jet of plasmoids have a Lorentz factor 
$\gamma_e\!\approx\!2\,\gamma^2$ in the blazar rest frame. Their 
maximal energy probably is similar to the observed knee energy of 
Galactic cosmic ray electrons (Dar and De R\'ujula 2008; De R\'ujula 
2019). This beam of high energy CR electrons, while propagating in 
the blazar halo and beyond, up-scatter x ray and gamma ray photons 
-whose origin is the blazar accretion disk and broad line region (Dar 
and Laor 1997), respectively- up to TeV energies through ICS in the 
Klein Nishina regime. These ICS photons have energies which extend 
effectively up to
\begin{equation}
h\nu_{max}\!\approx\! 2\gamma^2 m_e c^2\,,
\end{equation} 
in the jet direction of motion, and up to $\gamma^2 m_e c^2$ 
for a typical  viewing angle $\theta\!\sim\! 1/\gamma$
relative to the jet direction of motion, yielding
\begin{equation}
E_{p2}=h\nu_{p2}\!\approx\! \gamma^2 m_e c^2/2\,.
\end{equation}

\section{Universal Peaks Ratio In Blazars SED}
ICS of photons in the Klein Nishina regime by
CR electrons, which were accelerated by blazar jets, 
are very narrowly beamed along the  direction of motion
of the CR electrons.  Consequently, the scattered photons 
in the Thomson and Klein Nishina regimes share the
same beaming cone. Hence, the ratio of the locally observed 
peak energies of the ICS of photons
of the blazar x-ray/$\gamma$-ray halo (eq.3) and the 
peak energies of the ICS of  
CMB photons at the blazar redshift (eq.5), 
are expected to satisfy 
\begin{equation}
(1\!+\!z)\nu_{p2}/\nu_{p1}\!\approx\! 
m_e c^2/4\epsilon_p(CMB)\!\approx\! 1.88\times 10^8\, . 
\end{equation}
Note that although the broad band SED of double hump blazars,  
vary between blazars, and depends on their viewing angle, and on
time  due to blazar activity and deceleration of blazar jets
in the blazar's surrounding medium, Eq.6 implies that all these effects, 
cancel out to a good approximation in the double humps peaks ratio. This is
demonstrated first in Figures 1,2 for the best studied blazar, Mrk 421.  
Its  peaks ratio during the $\sim$ 13-day flaring event of Mrk 421 in March 2010 
(Aleksic et al. 2015) is compared in Figure 1 to its  average peaks ratio during the 
simultaneous multi-frequency observations from January 19 to June 1, 2009 
(Abdo et al. 2011b). 
\begin{figure}[]
\centering 
\epsfig{file=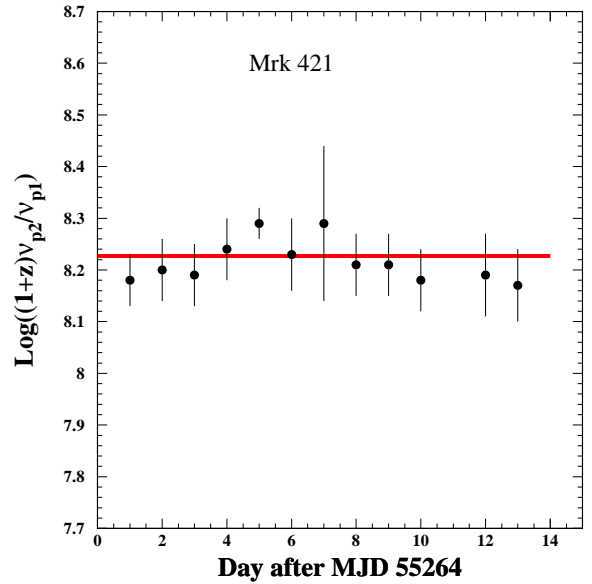,width=8.6cm,height=8.6cm}
\caption{The peaks ratio (eq.6) of the two broad 
humps in the SED of Mrk 421 as a function of time during 
its $\sim$ 13-day flaring event in March 2010 (Aleksic et al. 2015) 
and its weighted average value R=8.23 ($\chi^2/dof\!=\!0.78$).} 
\end{figure}

As can be seen from Figure 1, the ratio $R\!=\!log[(1\!+\!z)\nu_{p2}/\nu_{p1}]$ 
of the broad band peaks in the SED of the blazar Mrk 421 during $\sim$ 
13-day flaring event in March 2010 has not changed significantly 
as a function of time, and coincided within errors with its 
long term average value during the 
multi-frequency campaign from January 19 to June 1, 2009 (Abdo et al. 
2011b). Moreover, Figure 2  shows that, within errors, this peaks
ratio remained the same during the steady states, outburst, and flares 
of Mrk 421 in the 4.5 years period from August 2008 to February 2013 
(Bartoli et al. 2016). 
\begin{figure}[]
\centering
\epsfig{file=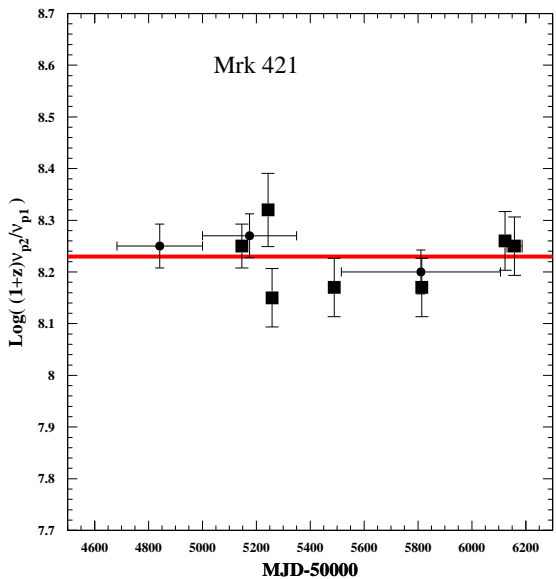,width=8.6cm,height=8.6cm}
\caption{The peaks ratio (eq.6) of the two broad
humps in the SED of Mrk 421 in flares (squares),
and steady states and outburst (circles) during 
the 4.5 years period from August 2008 to February 2013 (Bartoli et al. 2016).
The horizontal line is their weighted average value R=8.23 ($\chi^2/dof\!=\!1.13$).}  
\label{Fig2}
\end{figure}
The peak values of the broad humps in the SED of a 
representative sample of 12 double hump blazars with a 
known redshift and well determined peak frequencies 
of both humps obtained from multi-wavelength campaigns 
are listed in Table I. The $\nu_{p2}$ values have been extracted 
from published spectra which were corrected for absorption by 
extragalactic background light according to 
the model of Franceschini et al. (2008). 
\begin{table*}
\caption{SED peak ratio in representative sample
of double hump blazars}
\label{table:1}
\centering
\begin{tabular}{l l l l l}
\hline
\hline
~~~~Blazar~~~& $z$ & $log[(1\!+\!z)\nu_{p2}/\nu_{p1}]$ & MJD- & Data Summary \\ 
\hline
1ES 0229+200 & 0.1396 &~~~~ $\approx$ 8.40  & 55118- &  Aliu et al. 2014   \\ 
 
J0733.5+5153 & 0.065~ &~~~~  $\approx$ 8.38  & 58141- & Acciari et al. 2019b\\

Mrk 501~~~~~ & 0.0336 &~~~~  $\approx$ 8.34  & 56087- & Ahnen  et al. 2018\\  
             
Mrk 421~~~~~ & 0.0300 &~~~~  $\approx$ 8.23  & 54850- & Abdo et al. 2011b\\

1ES 1741+196 & 0.084~ &~~~~  $\approx$ 8.05  & 54940- & Abeysekara et al. 2016\\ 

1ES 1959+650 & 0.048~ &~~~~  $\approx$ 8.20  & 57547- & Acciari et al. 2020a \\

H 2356-309~  & 0.165~ &~~~~   $\approx$ 8.30  & 53534- & Aharonian et al. 2006\\

1ES 2344+54~~& 0.044~ &~~~~   $\approx$ 8.30  & 57611- & Acciari et al. 2020b \\ 

3C 66A~~~~~~~& 0.340~ &~~~~   $\approx$  8.30  & 54734- & Abdo et al. 2011a \\

1ES 1215+303 & 0.135~ &~~~~   $\approx$ 8.40  & 54682- & Valverde  et al. 2020\\

TXS 0506+056 & 0.5573 &~~~~   $\approx$ 8.22  & 58020- &  Aarsten et al. 2018 \\

3C 279~~~~~~ & 0.5362 &~~~~   $\approx$ 8.30  & 56741- & Larinov et al. 2020\\ 

\hline

\end{tabular}
\end{table*}
In Figure 3, $log((1\!+\!z)\nu_{p2})$ is plotted as a function 
of $log(\nu_{p1})$ for a representative sample of 12 blazars 
with a double hump SED which are listed in Table 1, 
the microquazar SS 433, and the gamma ray burst GRB190114C. 
The peak values of blazars were adopted from reported best 
fits/theoretical parametrizations of their SED extracted from 
multi-wavelength observation campaigns  cited in Table 1. 
The best fit line $log((1\!+\!z)\nu_{p2})\!=\!a\,log(\nu_{p1})\!+\!b$, 
to the double peak values of the blazars in Table I has 
yielded $a\!=\!1.0$ and $ b\!=\!8.38\,$ with $\chi^2/dof\!=\!0.16$.
\begin{figure}[]
\centering
\epsfig{file=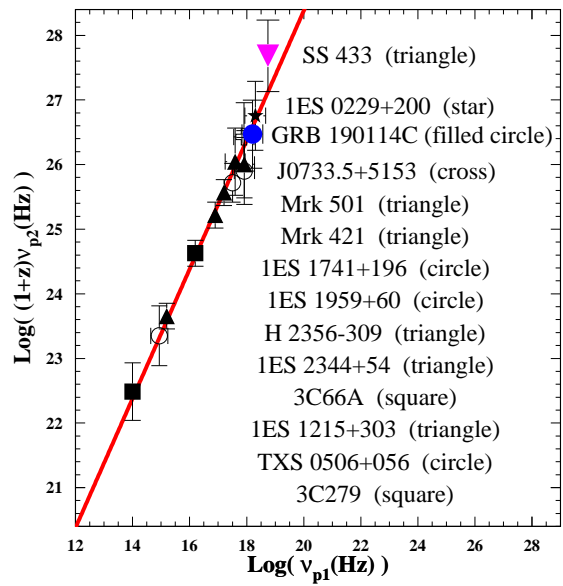,width=8.6cm,height=8.6cm}
\caption{$log[(1\!+\!z)\nu_{p2}]$ plotted as a function of 
$log(\nu_{p1})$ for a representative sample of 12 double hump blazars 
with known redshift and well measured peak values in simultaneous 
multi-wavelength observations. The line is the best fit 
line $log[(1\!+\!z)\nu_{p2}]\!=\!a\, log(\nu_{p1})\!+\! b$,
where  $a\!=\!1.0$, $b\!=\!8.38$, ($\chi^2/dof\!=\!0.16)\, .$} 
\end{figure}

\section{Double Hump Afterglow of GRBs ?} 
The detection of GeV photons following prompt emission pulses in GRBs 
by EGRET (Hurley et al. 1994) led to the suggestion (Dado \& Dar 2005) that 
the SED  of GRBs has a double hump 
structure similar to that observed in blazars. Recently multi-frequency 
observations of the ultrabright GRB 190114C at redshift $z=0.425$ 
(Selsing et al. 2019; Castro-Tirado, et al. 2019)
and its afterglow emission across 17 orders of magnitude in energy, from 
$5\times 10^{-6}$ to $10^{12}$ eV, found that its broad band 
SED, was double-peaked. Its high energy 
hump which was observed with the MAGIC telescope (Acciari et al. 
2019a)  peaked around TeV (after correcting for absorption by 
extragalactic background light), while the Fermi and Swift 
observations of GRB 190114C (Ajello et al. 2020) indicated a broad 
SED hump which peaked around 5.2 keV. The observed peaks ratio, 
$(1\!+\!z)E_{p2}/E_{p1}\!\approx\!2.74\times 10^8$, is consistent 
with that observed in blazars (eq. 6). This ratio in GRB 190114C
is also indicated in Figure 1.

 
\section{Double Hump SED Of Microqusars ?}
The Galactic microquasar SS 433 (e.g., Mirabel \& Rodriguez 1999 for a 
review of Galactic microquasars) is a binary system containing a compact 
object (either a stellar mass black hole or a neutron star) accreting 
matter from a supergiant star, which is overflowing its Roche lobe. 
Bipolar jet of plasmoids (cannonballs) which are ejected with a bulk 
velocity $\approx 0.26$c, extend from the binary 
perpendicular to the sightline and terminate in lobes 25 pc away, 
emit GeV-TeV gamma rays (Xing et al. 2019; Fang et al. 2020). 

If the electrons in these terminal lobes of SS 433 are Fermi 
accelerated there up to energies comparable to that of the knee 
energy of Galactic CR electrons, $E_{knee}(e)\!\sim\!2.3\!\pm\!0.7$ TeV 
(Ambrosi et al. 2017; De R\'ujula 2019), then ICS of CMB photons by 
such electrons inside/outside  the lobes of SS 433 yields high 
energy photons with a peak energy around
\begin{equation}
E_p\!\approx\![E_{knee}(e)/m_e]^2\,\epsilon_p(CMB) 
     \approx\! 14(+\!10, -\!4)\, {\rm  GeV}\, .  
\end{equation}
However, the hard x-ray continuum emission from SS 433 which 
peaks around $\sim 22$ keV, is consistent with being 
bremsstrhalung emission by the ambient electrons of the interstellar 
matter which enter the plasmoids of SS 433 with $\beta\!\approx\! 0.29$, 
(in the plasmoid rest frame) and decelerates there. 
This observed speed of the plasmoids 
implies a kinetic energy release 
$m_ec^2(\sqrt{1+\beta^2}-1)\!\approx\! 41$ keV within the 
plasmoids by swept in electron. This energy release may explain 
the hard x-ray emission observed by Suzaku (Kubota et al. 2010), 
INTEGRAL (Cherepashchuk et al. 2013), and NuSTAR (Middleton et 
al. 2019) X-ray satellites.  Hence the ratio of the broad band 
peaks of the SED of the microquasar SS 433 may have coincided by 
chance with the universal ratio observed in blazars.

\section{Conclusions and Implications}
The peak energies of the double humps of broad band spectral 
energy density of different blazars/GRBs are spread over a wide 
range and are time dependent. However,
their observed ratio (after correcting for pohoton absorption along 
their sightline) appears to satisfy, within observational errors, 
$(1\!+\!z)E_{p2}/E_{p1}\!\approx\!m_e c^2/4 \epsilon_p$. 
Such a universal ratio is obtained if both humps 
are produced by inverse Compton scattering (ICS); the lower hump by 
ICS of cosmic microwave background (CMB) photons in the Thomson 
regime, the high energy hump by ICS of x-rays and gamma rays in the 
Klein Nishina regime. 

Universal  peaks ratio in  blazars belonging to different 
classes in the BL Lac sequence, may reflect mainly different viewing 
angles and initial Lorentz factors of their jetted plasmoids, 
$\gamma(0)^2 \theta^2 \lsim 1$  of EHBL and HBL blazars and 
$\gamma(0)^2 \theta^2\!\gg\!1$ of IBL and LBL blazars of 
otherwise quite similar blazars. 

The energy spectrum of the cosmic ray electrons (CRe) in our Galaxy has 
a knee around $\approx\!2.3$ TeV (Ambrosi et al. 2017; De R\'ujula 
2019). Such a CRe knee was predicted by the cannonball (CB) model of
cosmic ray acceleration. In the CB model, the acceleration of 
ISM nuclei and electrons to high energies by
highly relativistic plasmoids ejected by blazars and gamma ray bursters 
yields energy spectra with a knee 
proportional to their mass (Dar \& De R\'ujula 2008), i.e.,
$E_{knee}(CRe)\!\approx (m_e/m_A)E_{knee}(m_A)$. If the 
knee energy of CRe is universal, that is, common to our Milky Way, 
external galaxies, and blazars, then the maximal peak energy 
of the observed humps in high-energy peaked blazars 
satisfy (after correcting for photon absorption)   
and the maximal peak energy of their low energy hump  satisfy,
\begin{equation}
{\rm max}\, E_{p1}\!\approx\!2 (E_{knee}(CRe)/m_ec^2)\epsilon_p\! 
\approx\!6\,{\rm keV},
\end{equation}
\begin{equation}
{\rm max}\, E_{p2}\!\approx\!E_{knee}(CRe)\!\approx\!2.3/(1\!+\!z) \,{\rm TeV}, 
\end{equation}
Such limits are consistent with the observed peak energies (after 
correcting for photon absorption along the sightline ) of EHBL blazars. 

The Lorentz factors of the plasmoids comprising the jets of blazars and 
GRBs have been estimated from measurements of their apparent 
superluminal velocity (Rees 1967) long time after ejection 
(in the blazar/GRB rest frame), {\bf assuming} that $\gamma(t)$, the 
Lorentz factor of the 
jet, and its viewing angle $\theta$ satisfy $\gamma \theta\!\approx\!1$. 
Typically, such estimates have yielded $\gamma(t)\!<\! 100 $, far below 
$\gamma(0)\!\approx\!1500$ needed to explain the knee energies of cosmic 
ray nuclei and electrons (Dar \& De R\'ujula 2008; De R\'ujula 2019). 
However, the observations of superluminal velocities of plasmoids 
launched by blazars/GRBs were carried out at relatively late times after 
launch (in the blazar/GRB rest frame), when the plasmoids have already 
decelerated considerably in the ISM of the host galaxy and satisfy 
$[\gamma(t)]^2\theta^2\!\ll\![\gamma(0)]^2\theta^2\!\approx\!1 $.
As long as $[\gamma(t)]^2\!\gg\!1$  and $\theta^2\!\ll\!1$, 
the apparent superluminal velocity of plasmoids satisfies 
\begin{equation}
V_{sl}\!=\!{\beta\,sin\theta\,c \over \!1-\!\beta\,cos\theta)} \!\approx 
\!{\gamma^2\theta^2 \over 1\!+\!\gamma^2\theta^2}{2c\over\theta}\,.
\end{equation}
The {\bf  assumption} $\gamma\theta\!\approx\! 1$ yields 
$\gamma\!=\!V_{sl}/c$.  However, highly relativistic 
plasmoids decelerate very fast with increasing time $t$ in the observer 
frame due to time abberation ($dt\!=\!(1\!+\!z)dt'/\gamma\delta$),  
yielding $\gamma^2\theta^2 \!\ll\! 1$ shortly after launch.
Consequently, as can be seen from eq.(10), most  measurements 
of the superluminal velocity of ejected plasmoids with
$\gamma(0)\theta \sim 1$  yield  
$V_{sl}(t)\!\ll V_{sl}(0)\!\sim\! \gamma(0)c$. 
 

{\bf Acknowledgement:} The authors thank A. De R\'ujula and an
anonymous referee for useful comments.

\end{document}